\documentclass[preprint,aps,12pt,showkeys,nofootinbib,tightenlines,superscriptaddress]{revtex4-1}

\usepackage{graphicx}
\usepackage{dcolumn}
\usepackage{bm}
\usepackage{amssymb}
\usepackage{amsfonts}
\usepackage{latexsym}
\usepackage{pdfsync}
\usepackage[T1]{fontenc} 
\usepackage{epsf,amsfonts}
\usepackage{epsfig}
\usepackage{amsthm}
\usepackage{amsmath}
\usepackage{color}
\usepackage{slashed}
\usepackage{mathrsfs}
\usepackage{float}
\usepackage{color}
\usepackage{esvect}
\usepackage{mathtools}
\usepackage{bbm}

\newcommand{\VEV}[1]{\left\langle #1\right\rangle}

\newcommand{\abs}[1]{\left| #1\right|}

\newcommand{\MeV}{\;\text{MeV}}
\newcommand{\GeV}{\;\text{GeV}}

\begin{document}

\title{Chiral transition and meson melting with finite chemical potential in an improved soft-wall AdS/QCD Model}

\author{Zhen Fang}
\email{zhenfang@hnu.edu.cn}
\affiliation{Department of Applied Physics, School of Physics and Electronics, Hunan University, Changsha 410082, China}

\author{Lin Zhang}
\email{zhanglin@itp.ac.cn}
\affiliation{School of Nuclear Science and Technology, University of Chinese Academy of Sciences, Beijing 100049, China}

\date{\today}

\begin{abstract}
We give a further study on the improved soft-wall AdS/QCD model with two flavors. The chiral transition behaviors are studied in the case of finite chemical potential, with the chiral phase diagram obtained at zero quark mass. The thermal spectral functions of the vector and axial-vector mesons are calculated, and the in-medium melting properties of the mesons are investigated. We find that the chiral transition behaviors and the meson melting properties at finite chemical potential can be qualitatively described by the improved soft-wall AdS/QCD model, except in the region of large chemical potential. The reason for these inadequate descriptions may be that the background geometry adopted in the model is not a dynamical one which is able to produce the QCD equation of state. To give a quantitative description for these low-energy phenomenologies, we shall consider a more consistent AdS/QCD model which treats the background fields and the chiral fields on the same footing.
\end{abstract}

\maketitle

\section{Introduction}\label{intro}

The issues on the strongly correlated quark-gluon-plasma (QGP) have been attracting a great deal of interest for the description of the low-energy properties of strong interaction. With the development of the collider physics and the powerful instruments for numerical computation, there have been many researches on the relevant properties of QGP. However, many puzzles remain to be solved to deep our understanding on the low-energy regime of quantum chromodynamics (QCD), which has motivated the development of various nonperturbative methods. Lattice QCD is a powerful tool from the first principle to look into the low-energy properties of QCD, yet this method can hardly address the issues with finite chemical potential due to the sign problem.

The anti-de Sitter/conformal field theory (AdS/CFT) correspondence builds a bridge between the type IIB superstring theory in AdS$_5\times S^5$ and the $\mathcal{N}=4$ super Yang-Mills theory on the boundary \cite{Maldacena:1997re,Gubser:1998bc,Witten:1998qj}, which provides us with another powerful method to treat the low-energy problems of QCD. In the holographic framework, we make use of a weakly coupled dual gravitational theory in the bulk to describe the strong-coupling properties of QCD at low energy, which is usually called AdS/QCD. In order to find a consistent quantitative description for the low-energy QCD, a large amount of works has been done in this field, ranging from the low-energy hadron physics to the thermodynamical properties of QCD \cite{Kruczenski:2003uq,Sakai:2004cn,Sakai:2005yt,deTeramond:2005su,DaRold:2005mxj,Erlich:2005qh,Karch:2006pv,Csaki:2006ji,Cherman:2008eh,Gherghetta:2009ac,Kelley:2010mu,Sui:2009xe,Sui:2010ay,Cui:2013xva,Fujita:2009wc,Fujita:2009ca,Colangelo:2009ra,Cui:2014oba,Li:2012ay,Li:2013oda,Shuryak:2004cy,Brodsky:2014yha,Tannenbaum:2006ch,Policastro:2001yc,Cai:2009zv,Cai:2008ph,Sin:2004yx,Shuryak:2005ia,Nastase:2005rp,Nakamura:2006ih,Sin:2006pv,Janik:2005zt,Li:2014hja,Li:2014dsa,Fang:2016uer,Fang:2016dqm,Lv:2018wfq,Attems:2016ugt,Attems:2017zam}.

There are mainly two approaches by which AdS/QCD is implemented. The top-down approach can give qualitative descriptions for certain low-energy hadron properties through the string-theoretic constructions with D-branes \cite{Kruczenski:2003uq,Sakai:2004cn,Sakai:2005yt}, while the bottom-up approach focuses on the characteristic features of low-energy QCD to construct dual models which are able to describe various low-energy phenomenologies of QCD \cite{deTeramond:2005su,DaRold:2005mxj,Erlich:2005qh,Karch:2006pv,Csaki:2006ji}. The hard-wall and soft-wall models are the typical ones in the bottom-up approach. The hard-wall model uses a sharp IR cut-off along the fifth dimension to realize the confinement \cite{DaRold:2005mxj,Erlich:2005qh}. Although with a proper chiral symmetry breaking, the linear confinement property of the spectrum cannot be obtained from the hard-wall model. In the soft-wall model, the hard cut-off is replaced by an exponential depressing term of a dilaton field in order to produce the Regge trajectory of the spectrum \cite{Karch:2006pv}. Nevertheless, the original soft-wall model has no spontaneous chiral symmetry breaking \cite{Colangelo:2011sr}, which is a fundamental feature of low-energy QCD.

The issue of QCD phase transition has been addressed in many holographic works \cite{Herzog:2006ra,BallonBayona:2007vp,Cai:2007zw,Kim:2007em,Andreev:2009zk,Colangelo:2010pe, Gubser:2008yx,Gubser:2008ny,Gubser:2008sz,DeWolfe:2010he,Gursoy:2007cb,Gursoy:2007er,Gursoy:2008bu,Gursoy:2008za,Rougemont:2015wca,Li:2011hp,Cai:2012xh,He:2013qq,Yaresko:2013tia,Finazzo:2014zga,Yang:2014bqa,Fang:2015ytf,Evans:2016jzo,Mamo:2016xco,Dudal:2016joz,Dudal:2018rki,Ballon-Bayona:2017dvv,Li:2017ple,Zollner:2018uep,ChenXun:2019zjc}. The deconfining phase transition was argued to be a first-order Hawking-Page transition between the thermal AdS and the AdS/Schwarzschild black-hole geometries, and the deconfinement temperature in the original soft-wall model is close to the lattice result \cite{Herzog:2006ra}. As the chiral condensate is naturally incorporated in the bifundemental scalar field of the soft-wall model, the chiral transition properties can also be studied in this type of models \cite{Colangelo:2011sr}. It has been shown that the chiral transition behaviors obtained from some modified soft-wall models are consistent with the standard scenario of QCD phase transition for both the two-flavor and the $2+1$ flavor cases \cite{Chelabi:2015cwn,Chelabi:2015gpc,Fang:2016cnt,Fang:2016nfj,Li:2016smq,Bartz:2016ufc,Bartz:2017jku,Chen:2018msc,Fang:2018vkp,Fang:2018axm,Fang:2019lmd}. Another important aspect of low-energy QCD is about the in-medium properties of hadrons. The thermal spectral functions for both mesons and glueballs have been studied in the soft-wall model, and the meson melting properties such as the in-medium mass shift have also been investigated \cite{Fujita:2009wc,Fujita:2009ca,Colangelo:2009ra,Cui:2014oba}.

In Ref. \cite{Fang:2016nfj}, we proposed a simply improved soft-wall AdS/QCD model with a running bulk scalar mass and a quartic action term of the scalar field in the two-flavor case. The light meson spectra and the chiral transition behaviors obtained from this model are consistent with the experiment data or the lattice results. Here we will give a further study on this improved soft-wall model in the case of finite chemical potential. The chiral transition behaviors and the melting spectral functions of the vector and axial-vector mesons will be investigated. Following the previous studies, we use the AdS/Reissner-N\"{o}rdstrom (AdS/RN) black hole to describe the temperature and chemical potential effects. 

The paper is organized as follows. In Sec. \ref{sec-model}, we outline the improved soft-wall AdS/QCD model. In Sec. \ref{sec-chiral}, we investigate the chemical potential effects on the chiral transition behaviors and obtain the chiral phase diagram in the chiral limit. In Sec. \ref{spectr-func}, we consider the in-medium properties of the mesons in the improved soft-wall model. The spectral functions of the vector and axial-vector mesons will be calculated in the case of finite chemical potential, and the meson melting properties will be studied. In Sec. \ref{conclusion}, we summarize the work and conclude with some remarks.

\section{The improved soft-wall AdS/QCD model}\label{sec-model}

\subsection{The action of the model}\label{model-review}

We first give a brief review of the improved soft-wall AdS/QCD model with two flavors which is proposed in Ref. \cite{Fang:2016nfj}. The bulk action of this model can be written as
\begin{align}\label{model-action}
S_M =\int d^{5}x\sqrt{g}e^{-\Phi(z)}\mathrm{Tr}\left\{|DX|^{2}-m_5^2(z)|X|^{2}-\lambda |X|^{4} -\frac{1}{4g_{5}^2}(F_{L}^2+F_{R}^2)\right\},
\end{align}
where the covariant derivative of the bulk scalar field is $D^MX=\partial^MX -i A_L^MX +i X A_R^M$, and the field strengths $F_{L,R}^{MN} =\partial^MA_{L,R}^N-\partial^NA_{L,R}^M-i[A_{L,R}^M,A_{L,R}^N]$ with the gauge fields $A_{L,R}^M$ in the representations of $\mathrm{SU}(2)_{L,R}$. The gauge coupling $g_5^2=\frac{12\pi^2}{N_c}$ is obtained by comparing the holographic calculation for the two-point correlation of the vector current $J_{\mu}^a=\bar{q}\gamma_{\mu}t^aq$ with the QCD result \cite{Erlich:2005qh}. The dilaton field takes the simplest form $\Phi(z)= \mu_g^2\,z^2$ to reproduce the Regge trajectory of the light meson spectrum \cite{Karch:2006pv}, and the quartic term of the bulk scalar field is crucial to generate spontaneous chiral symmetry breaking. One essential feature of this improved soft-wall model is a running bulk scalar mass with the form $m_5^2(z)=-3-\mu_c^2z^2$, which is necessary for the realization of right chiral transition behaviors.

In Ref. \cite{Fang:2016nfj}, we calculate the mass spectra of the pseudoscalar, scalar, vector and axial-vector mesons which are compared with the experimental results. The chiral transition behaviors with zero chemical potential have also been investigated, and the model results are consistent with the lattice QCD for the two-flavor case \cite{Laermann:2003cv,Kanaya:2010qd}.

\subsection{The background geometry}\label{sec-background}

As the interests of us here are the in-medium spectral properties of the light mesons and the chiral transition behaviors with finite chemical potential, we will take the AdS/RN black hole as the bulk background of the improved soft-wall model. This bulk geometry is the solution of the Einstein-Maxwell system with the prescription of the $U(1)$ gauge field: $A_{i}=A_{z}=0$, $A_{0}=A_{0}(z)$ and the metric ansatz:
\begin{align}\label{metric}
&ds^2=e^{2A(z)}\left( f(z)\,dt^{2}-{d{x^i}}^2-\frac{dz^2}{f(z)} \right),
\nonumber \\
&A(z)=-\log{\frac{z}{L}},
\end{align}
which yields (we will set the AdS radius $L=1$ below)
\begin{align}\label{func_in_metric}
&A_0(z)=\mu -\kappa q z^2,     \nonumber \\
&f(z)=1 -(1+Q^2)\left(\frac{z}{z_h}\right)^4+Q^2\left(\frac{z}{z_h}\right)^6,
\end{align}
where $Q=q_B z_h^3$ with $q_B$ denoting the charge of the black hole and $z_h$ being the location of the event horizon, and $\kappa$ is a dimensionless constant which scales as $\sqrt{N_c}$ (we will set $\kappa=1$ for simplicity) \cite{Colangelo:2011sr}. 

The chemical potential $\mu$ is determined by the condition $A(z_h)=0$ as
\begin{align}\label{chem-potent}
\mu=\kappa\frac{Q}{z_h}=\frac{Q}{z_h},
\end{align}
and the Hawking temperature is defined as
\begin{align}\label{Hawking-T}
T =\frac{1}{4\pi}\abs{\frac{df}{dz}}_{z=z_h} =\frac{1}{\pi z_h}\left( 1-\frac{\mu^2 z_h^2}{2} \right)
\end{align}
with $0<\mu z_h<\sqrt{2}$.

\section{Chiral transition with finite chemical potential}\label{sec-chiral}

\subsection{The EOM of the scalar VEV}\label{sec-scalar-vev}

In the soft-wall AdS/QCD model, the chiral condensate, as the (approximate) order parameter of chiral transition, is implicitly incorporated in the UV expansion of the vacuum expectation value (VEV) of the bulk scalar field $\VEV{X}$. Thus, we first consider the scalar VEV part of the improved soft-wall model in order to study the chiral transition behaviors at finite $\mu$. By convention, we define $\VEV{X} =\frac{\chi(z)}{2}I_2$ with $I_2$ denoting the $2\times2$ identity matrix. The action of the scalar VEV $\chi(z)$ can be read from the bulk action (\ref{model-action}) as 
\begin{align}\label{chi-act}
S_{\chi} = \int d^5x\sqrt{g}e^{-\Phi}\left[\frac{1}{2}g^{zz}(\partial_z\chi)^2 -\frac{1}{2}m_5^2(z)\chi^{2} -\frac{\lambda}{8}\chi^{4}\right],
\end{align}
from which the equation of motion (EOM) of $\chi(z)$ can be derived as
\begin{align}\label{eom-chi}
\chi'' +\left(3A'-\Phi'+\frac{f'}{f}\right)\chi' -\frac{e^{2A}}{f}\left(m_5^2\,\chi +\frac{\lambda}{2}\chi^3\right) =0.
\end{align}

According to the AdS/CFT dictionary \cite{Erlich:2005qh}, the UV asymptotic form of $\chi(z)$ can be obtained from Eq. (\ref{eom-chi}) as
\begin{align}\label{UV-chi}
\chi(z\sim 0) = m_q\zeta z +\frac{\sigma}{\zeta}z^3 +\frac{1}{4}\,m_q\zeta z^3\left(m_q^2\zeta^2\lambda +4\mu_g^2 -2\mu_c^2\right)\log z +\cdots ,
\end{align}
where $m_q$ and $\sigma$ represent the current quark mass and the chiral condensate respectively, and $\zeta=\frac{\sqrt{N_c}}{2\pi}$ is a normalization constant which is necessary for the correct $N_c$ scaling behavior of $m_q$ and $\sigma$ \cite{Cherman:2008eh}. There is also a natural IR boundary condition for Eq. (\ref{eom-chi}) in order to make $\chi(z)$ regular near the horizon,
\begin{align} \label{chi-eom-BC}
\left[f'\chi' -e^{2A}\left(m_5^2\,\chi +\frac{\lambda}{2}\chi^3\right)\right]_{z=z_h} =0 .
\end{align}

In the numerical calculation, we will take advantage of the identity $\chi'(0) =m_q\zeta$ as the UV boundary condition, which follows from the UV asymptotic form (\ref{UV-chi}). With the given boundary conditions, we can solve Eq. (\ref{eom-chi}) numerically and extract the value of the chiral condensate $\sigma$ as functions of $\mu$ and $T$ from the UV expansion of $\chi(z)$. Thus, we can investigate the chiral transition behaviors with finite chemical potential.

\subsection{Numerical results}\label{sec-scalar-vev}

Now we conduct the numerical calculation to solve Eq. (\ref{eom-chi}) with the following set of parameter values: $m_q =3.22 \MeV$, $\mu_g =440 \MeV$, $\mu_c =1450 \MeV$ and $\lambda =80$, which has been shown to provide a good description for the mass spectra of the pseudoscalar, vector and axial-vector mesons and also the correct chiral transition behaviors for the two-flavor case with zero chemical potential.

\begin{figure}
\begin{center}
\includegraphics[width=75mm,clip=true,keepaspectratio=true]{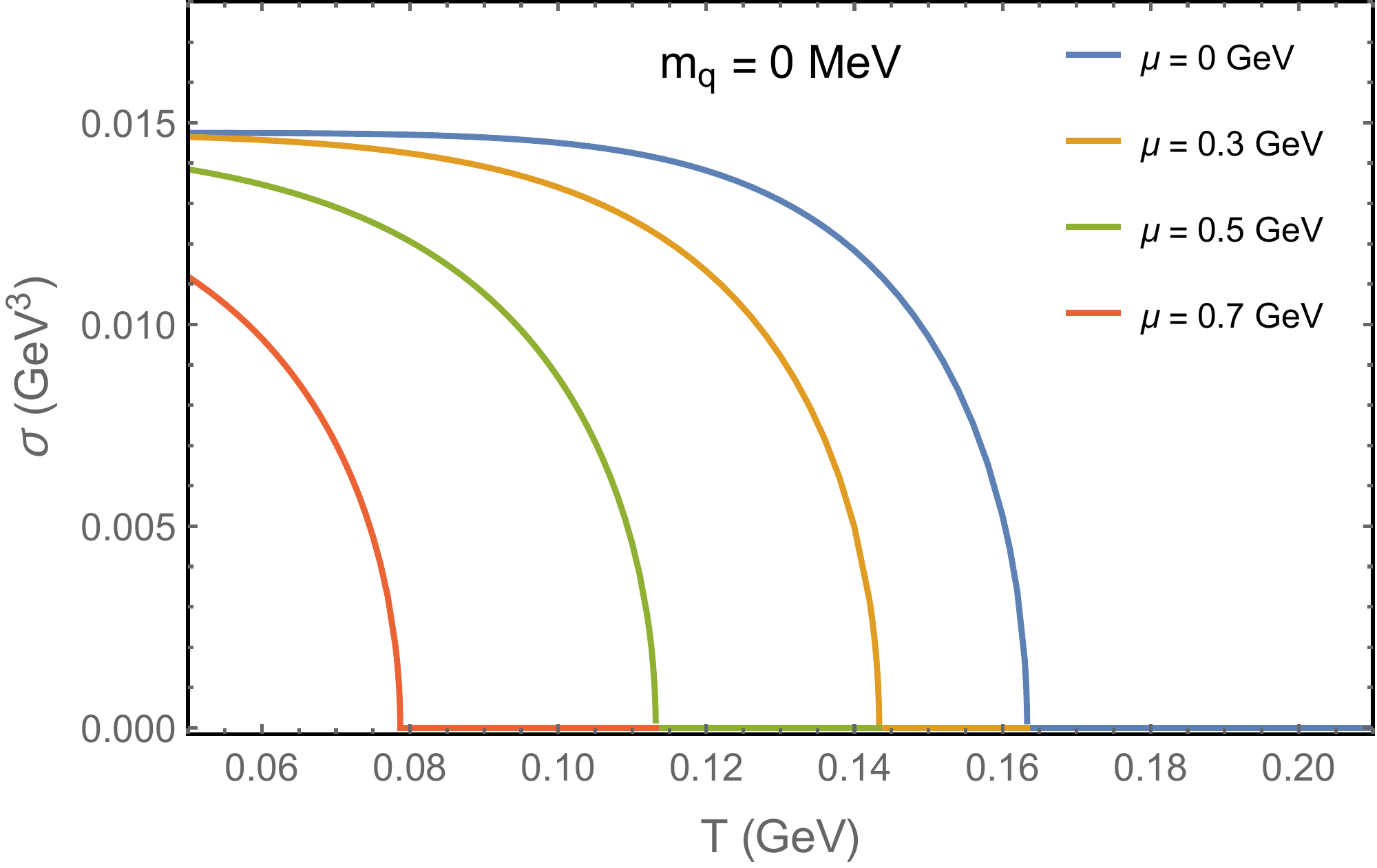} 
\vskip 0.3cm  \hskip 0.2 cm
\includegraphics[width=77mm,clip=true,keepaspectratio=true]{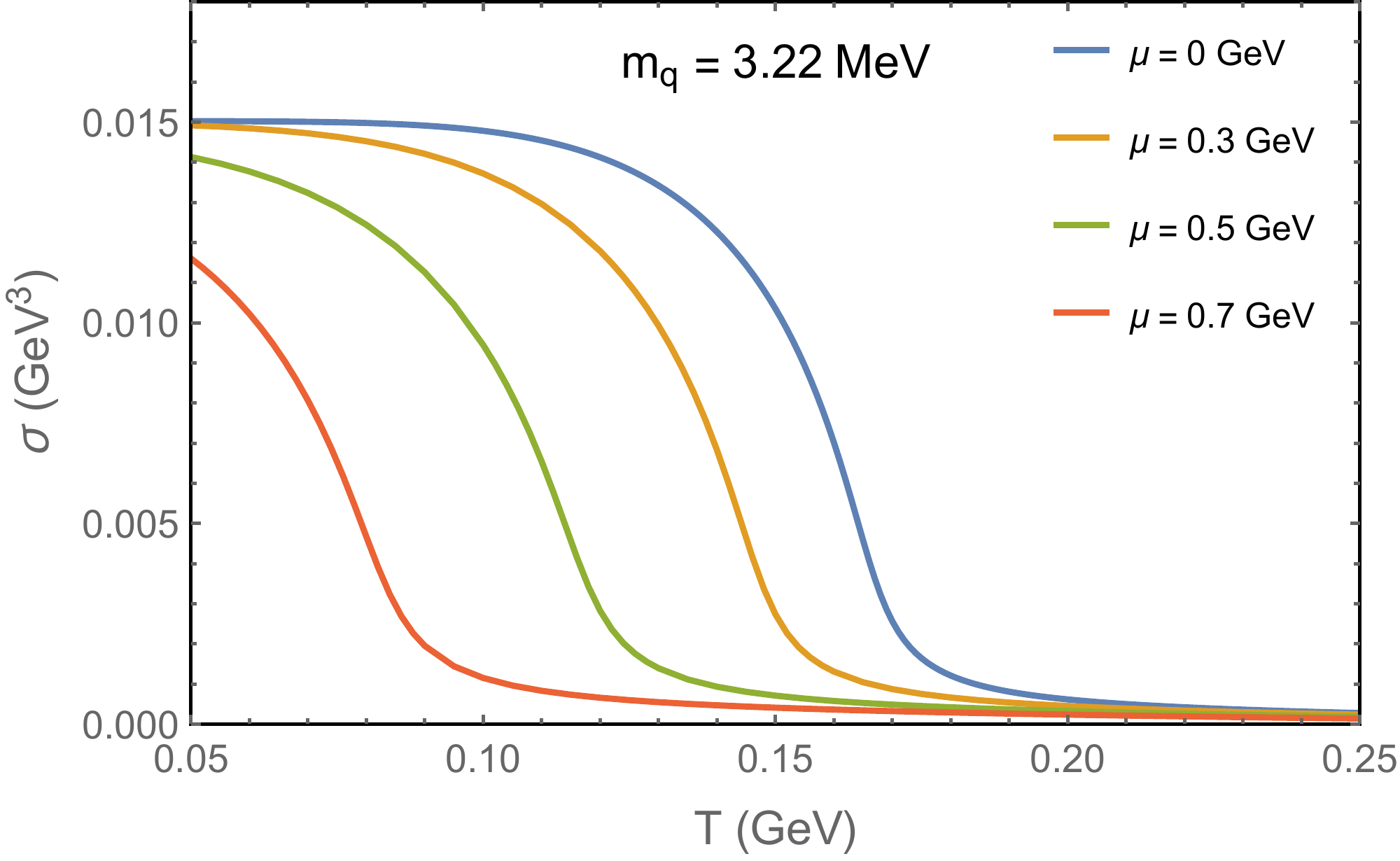}
\vskip -0.8cm \hskip 0.7 cm
\end{center}
\caption{The chiral transition behaviors with the temperature $T$ at four different chemical potentials: $\mu=0, 0.3, 0.5, 0.7 \GeV$ in the case of $m_q=0 \MeV$ (upper panel) and $m_q=3.22 \MeV$ (lower panel).}
\label{fig-sigma-T}
\end{figure}

First, we investigate the transition behaviors of the chiral condensate $\sigma$ with the temperature $T$ at four different chemical potentials in the case of $m_q=0 \MeV$ and $m_q=3.22 \MeV$ respectively. The numerical results obtained from the model are shown in Fig. \ref{fig-sigma-T}, where we can see that a second-order phase transition happens in the chiral limit, while it becomes a crossover transition when there is a nonzero quark mass. These chiral transition behaviors are consistent with the lattice QCD indications \cite{Kanaya:2010qd}. Furthermore, the (pseudo-)critical temperature $T_c$ deceases as expected with the increase of $\mu$. We also find that the order of chiral transition will not change with the chemical potential $\mu$ in the improved soft-wall AdS/QCD model with two flavors.

\begin{figure}
\begin{center}
\includegraphics[width=75mm,clip=true,keepaspectratio=true]{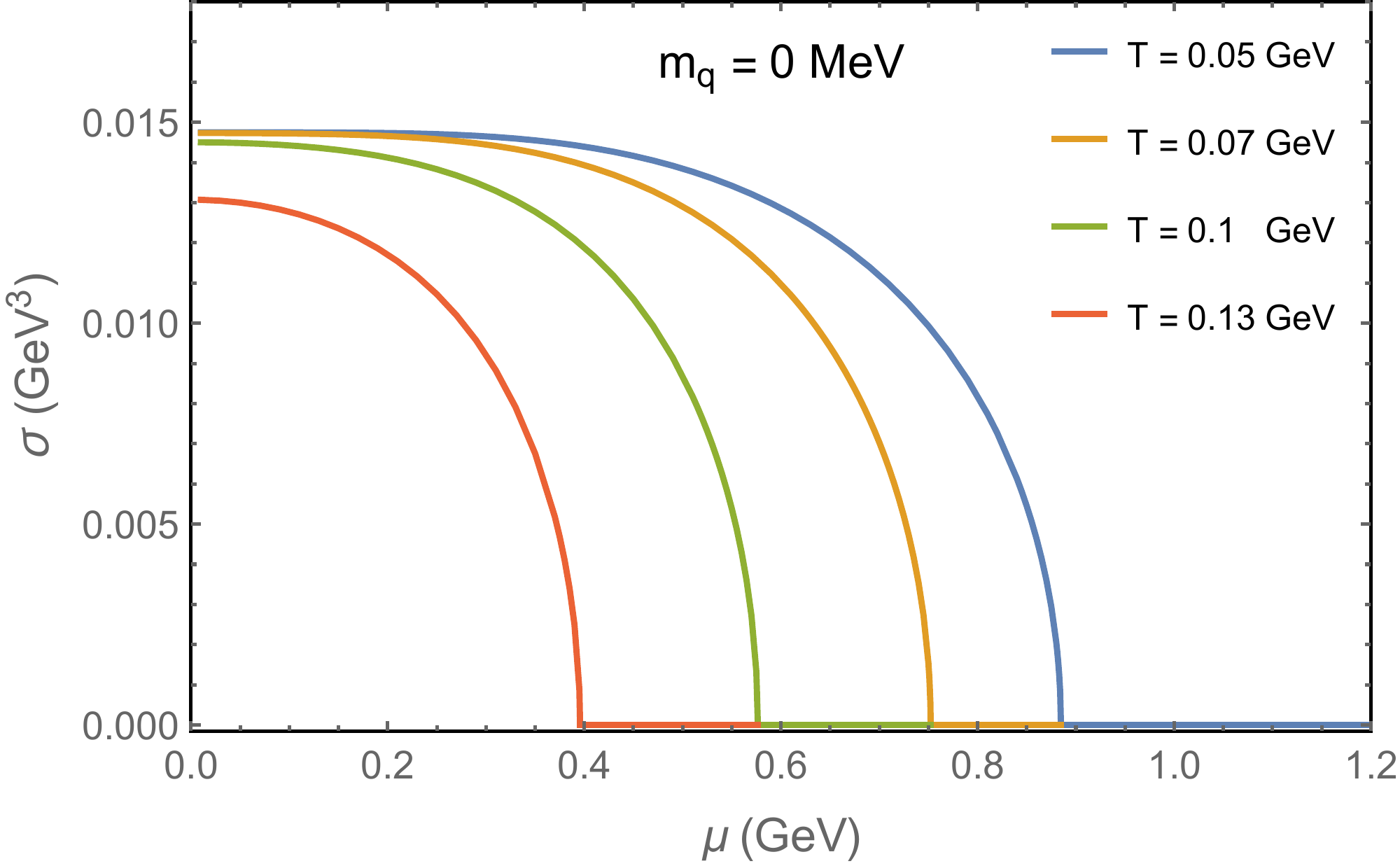}
\vskip 0.3cm  \hskip -0.1 cm
\includegraphics[width=73mm,clip=true,keepaspectratio=true]{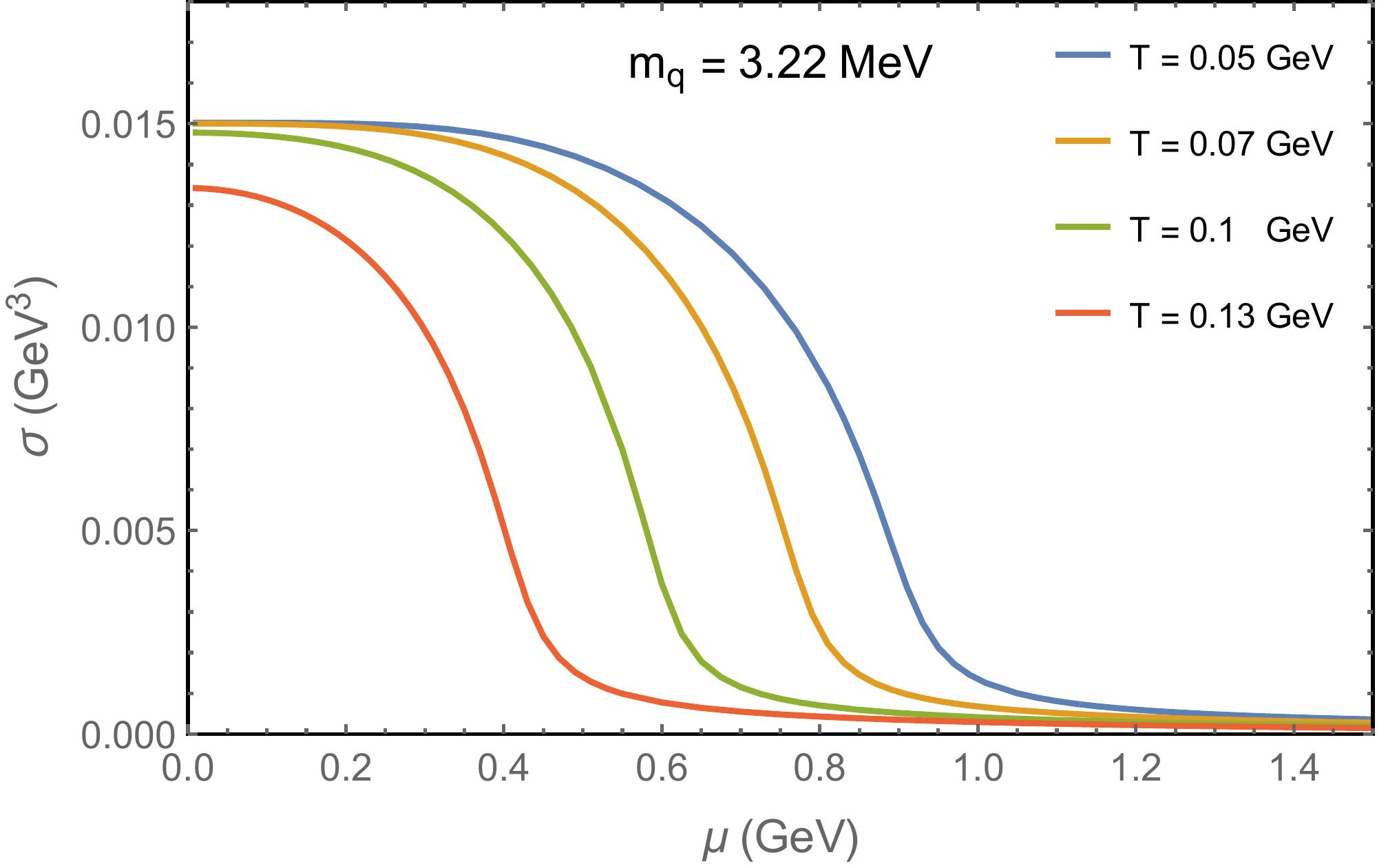}
\vskip -0.8cm \hskip 0.7 cm
\end{center}
\caption{The chiral transition behaviors with the chemical potential $\mu$ at four different temperatures: $T=0.05,  0.07, 0.1, 0.13 \GeV$ in the case of $m_q=0 \MeV$ (upper panel) and $m_q=3.22 \MeV$ (lower panel).}
\label{fig-sigma-mu}
\end{figure}

\begin{figure}
\centering
\includegraphics[width=73mm,clip=true,keepaspectratio=true]{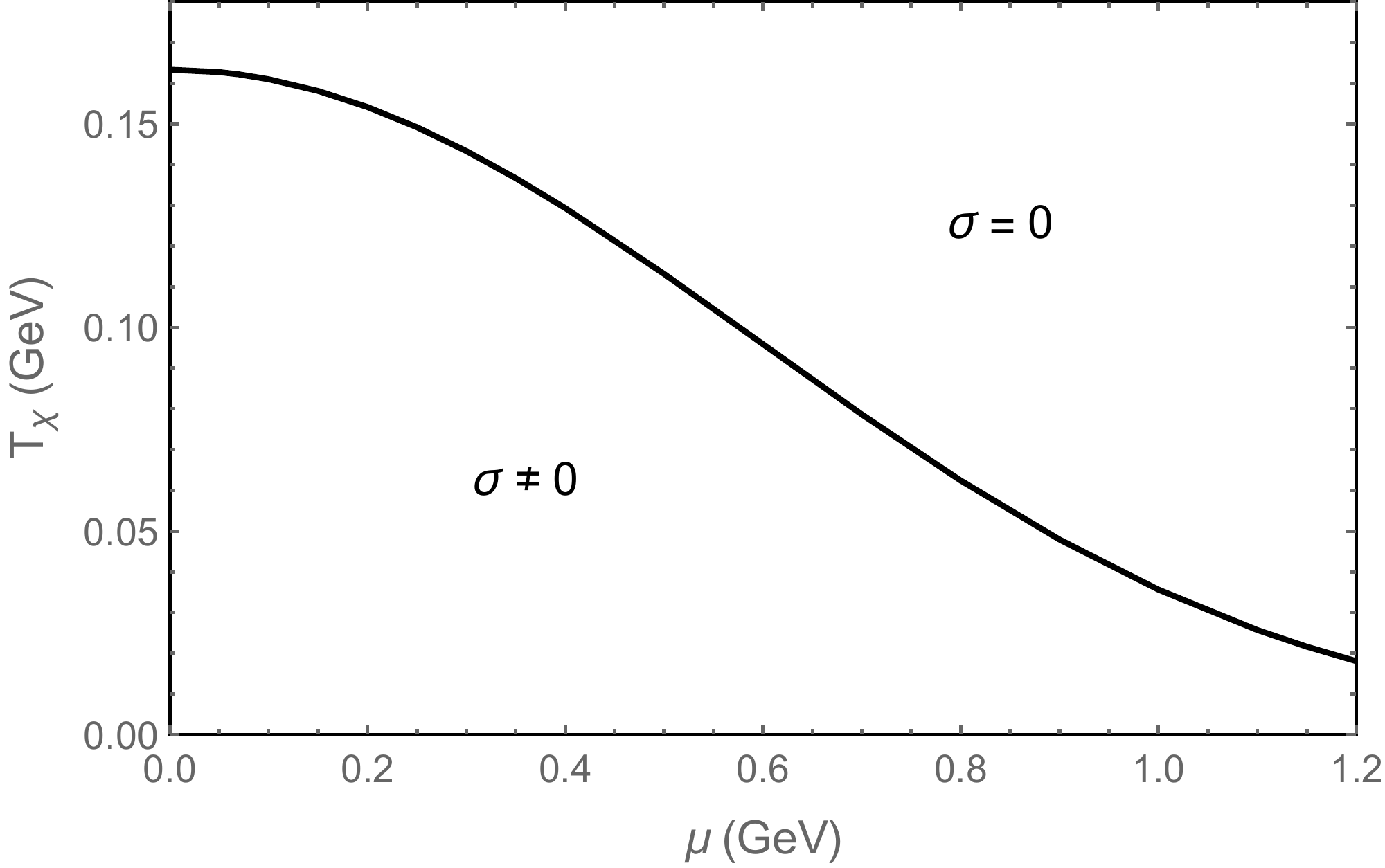}
\caption{The chiral phase diagram in the $\mu-T$ plane for the case of $m_q=0 \MeV$.} 
\label{fig-T-mu-chi}
\end{figure}

We then investigate the chiral transition behaviors with the chemical potential $\mu$ at four different temperatures, which are shown in Fig. \ref{fig-sigma-mu}. It can be seen that the dependences of $\sigma$ with $\mu$ have the same qualitative behaviors as those shown in Fig. \ref{fig-sigma-T}. We plot the chiral phase diagram in the $\mu-T$ plane for the case of zero quark mass in Fig. \ref{fig-T-mu-chi}, where the critical temperature $T_{\chi}$ is defined as the one at which $\sigma$ declines to zero. As in the $2+1$ flavor case \cite{Fang:2018axm}, we find that the second-order critical line descends too slowly at large $\mu$, which indicates that this improved soft-wall model cannot be applied in the far regions of large chemical potential. Actually, as we know, the AdS/RN black hole adopted here is holographically related to a conformal gauge theory, whose properties are rather different from those of QCD without conformal invariance \cite{Gubser:2008yx}. To introduce the temperature and chemical potential effects in a more reasonable way, we shall consider a dynamical background solved from an Einstein-Maxwell-dilaton system which breaks conformal invariance \cite{DeWolfe:2010he}. 

\section{Spectral function of the (axial-)vector meson}\label{spectr-func}

Now we consider the meson melting properties by looking into the spectral functions of the vector and axial-vector mesons. The gauge field sector of the bulk action (\ref{model-action}) can be rewritten by the vector field and the axial-vector field with the definitions $V=\frac{1}{2}(A_L+A_R)$ and $A=\frac{1}{2}(A_L-A_R)$,
\begin{align}\label{V-A-action}
S_{va}  =-\frac{1}{2g_{5}^2}\int d^{5}x\sqrt{g}e^{-\Phi(z)}\mathrm{Tr}\{F_V^2+F_A^2\},
\end{align}
where the field strengths of the vector and axial-vector gauge fields have the form: $F_{V}^{MN} =\partial^MV^N-\partial^NV^M-i[V^M,V^N]-i[A^M,A^N]$ and $F_{A}^{MN} =\partial^MA^N -\partial^NA^M -i[V^M,A^N] -i[A^M,V^N]$.


\subsection{The vector meson}

We first consider the in-medium spectral properties of the vector meson, whose EOM can be derived from the variation of the action (\ref{V-A-action}) with respect to $V_{\mu}$ in the $V_z=0$ gauge,
\begin{align}\label{vec-eom1}
\partial_z\left(e^{-\Phi}\sqrt{g}\,g^{zz}g^{\mu\lambda}\partial_{z}V_\lambda\right) +e^{-\Phi}\sqrt{g}\,g^{\mu\lambda}g^{\nu\rho}\partial_{\nu}\partial_{\rho}V_{\lambda} =0.
\end{align}
By performing the Fourier transformation $V_{\mu}(x,z) =\int d^4x e^{ip\cdot x}V(p,z)\mathcal{V}^0_{\mu}(p)$ in terms of the source $\mathcal{V}^0_{\mu}(p)$ and the bulk-to-boundary propagator $V(p,z)$, and substituting the metric ansatz (\ref{metric}) into Eq. (\ref{vec-eom1}), we obtain the EOM of $V(p,z)$ for the spatial part of the vector field,
\begin{align}\label{vec-eom2}
\partial_z\left(e^{A -\Phi}f\partial_zV(p,z)\right) +e^{A -\Phi}\left(\frac{1}{f}\,\omega^2 -q^2 \right)V(p,z) =0
\end{align}
with $p^{\mu}=(\omega, q^1, q^2, q^3)$ and $q^2=q_1^2+q_2^2+q_3^2$.

To handle the above EOM conveniently, we apply the following transformation of the variables (note that we hide the momentum-dependence in $v(u)$ for simplicity):
\begin{align}\label{vec-coord1}
z \rightarrow u\,z_h, \quad   V(p,z) \rightarrow v(u), \qquad 0<u<1,
\end{align}
in terms of which Eq. (\ref{vec-eom2}) can be cast into the form:
\begin{align}\label{eom-v-u-vec}
v''(u) +k_1(u) v'(u) +k_0(u) v(u) =0
\end{align}
with
\begin{align}\label{eom-v-u-coef}
k_1(u) &=\frac{1}{u\left(1-u^2\right)\left(1+u^2-Q^2 u^4\right)}\Big[-2Q^2\mu_g^2 z_h^2\,u^8 +\left(2Q^2\mu_g^2 z_h^2 +2\mu_g^2 z_h^2 +5Q^2\right)u^6
\nonumber \\
&\quad -3\left(Q^2 +1\right)u^4 -2\mu_g^2 z_h^2\,u^2 -1\Big] ,        \nonumber\\
k_0(u) &=\frac{z_h^2}{\left(u^2-1\right)^2\left(1+u^2-Q^2 u^4\right)^2}\Big[\omega^2 -q^2\left(1 -\left(Q^2 +1\right) u^4 +Q^2 u^6\right)\Big] .
\end{align}

The asymptotic solution of Eq. (\ref{eom-v-u-vec}) near the horizon can be obtained as
\begin{align}\label{asy-vec-horizon}
v(u\sim 1) = c_{+}\,\phi_{+}(u) +c_{-}\,\phi_{-}(u)
\end{align}
with
\begin{align}\label{in-out-sol}
\phi_{\pm}(u) &=\left(1-u\right)^{\pm\mathrm{i}\frac{z_h\omega}{2\left(2-Q^2\right)}} ,
\end{align}
where $\phi_{+}$ represents the out-coming solution and $\phi_{-}$ represents the in-falling solution.
The UV asymptotic form of $v(u)$ can also be obtained by the Frobenius method as
\begin{align}\label{asy-vec-boundary}
v(u\sim 0) = A\left(1 +c_{l2}u^2\log{u} +c_{l4}u^4\log{u} +c_{4}u^4 +\cdots\right) +B\left(u^2+d_4 u^4 +\cdots \right)
\end{align}
with
\begin{align}\label{uv-coef-vec}
c_{l2} &= \frac{1}{2} z_h^2\left(q^2 -\omega^2\right),      \nonumber\\
c_{l4} &= \frac{1}{16} z_h^4\left(q^2 -\omega^2\right)\left(4\mu_g^2 +q^{2} -\omega^2\right),      \nonumber\\
c_4 &= -\frac{1}{64} z_h^4\left(q^2 -\omega^2\right)\left(4\mu_g^2 +3q^{2} -3\omega^2\right),      \nonumber\\
d_4 &= \frac{1}{8} z_h^2\left(4\mu_g^2 +q^{2} -\omega^2\right),
\end{align}
where $A, B$ are two arbitrary complex coefficients. In the numerical procedure, we set the coefficient $A=1$ to fix the overall constant in $v(u)$. The coefficient $B$ can then be uniquely determined once the IR boundary condition of $v(u)$ is specified.

Previous works have provided a detailed analysis on the holographic computation of the Green's function in the Minkowskian space-time \cite{Son:2002sd,Policastro:2002se}. According to Ref. \cite{Son:2002sd}, the retarded Green's function corresponds to the in-falling asymptotic solution of $v(u)$ with $c_{+}=0$ in (\ref{asy-vec-horizon}) near the horizon. By imposing this IR boundary condition, we can solve Eq. (\ref{eom-v-u-vec}) numerically, and extract the coefficient $B(\omega, q)$ as a function of $\omega$ and $q$. Following the prescription in Ref. \cite{Fujita:2009ca}, we relate the retarded Green's function $D^{R}(\omega,q)$ to the UV behavior of $v(u)$ as
\begin{align}\label{green-uv}
D^{R}(\omega,q) &=-\frac{C}{z_h^2}\lim_{u\to\epsilon}\left(\frac{1}{u}v^{*}v^{\prime}\right)       \nonumber\\
&=-\frac{2C}{z_h^2}\left[B(\omega, q) +\frac{1}{2}c_{l2} +c_{l2}\log\epsilon\right]
\end{align}
with the constant $C=N_c^2/(64\pi^2 L)$. The spectral function of the vector meson can then be obtained by its definition as
\begin{align}
\rho_v(\omega,q) =-\frac{1}{\pi}\,\mathrm{Im}D^R(\omega,q) =\frac{2C}{\pi z_h^2}\,\mathrm{Im}B(\omega, q).
\end{align}

\begin{figure}
\begin{center}
\hskip -0.2 cm
\includegraphics[width=71mm,clip=true,keepaspectratio=true]{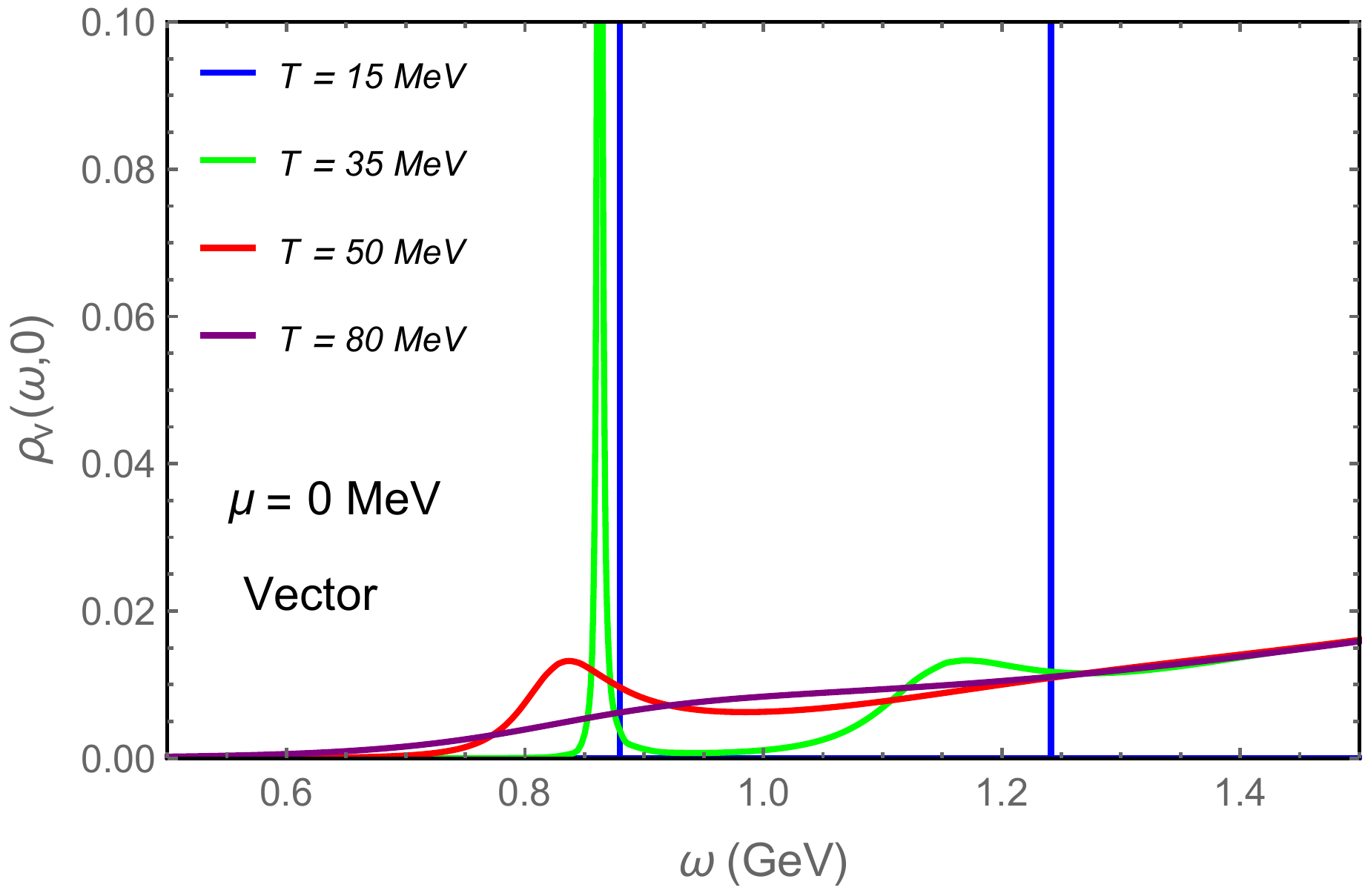}
\vskip 0.3cm  \hskip -0.1 cm
\includegraphics[width=72mm,clip=true,keepaspectratio=true]{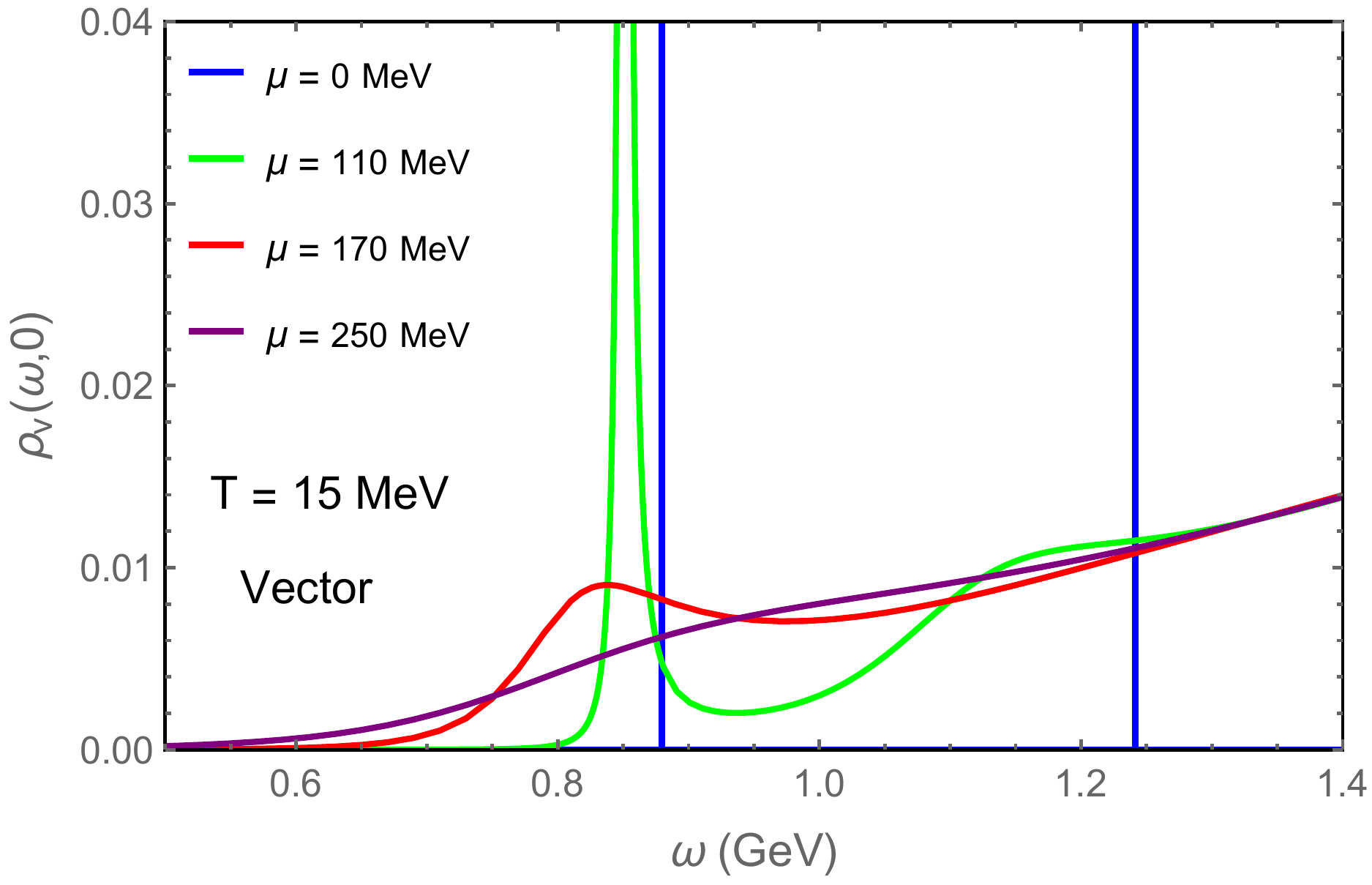}
\vskip -0.8cm \hskip -0.1cm
\end{center}
\caption{The spectral functions of the vector meson $\rho_v(\omega,0)$. The upper panel shows $\rho_v(\omega,0)$ at four different temperatures in the $\mu=0$ case, while the lower panel shows $\rho_v(\omega,0)$ at four different chemical potentials for a fixed temperature $T=15\MeV$.}
\label{spectr-V}
\end{figure}

The numerical calculations for the spectral function of the vector meson $\rho_v(\omega,q)$ are presented in Fig. \ref{spectr-V}, where we plot $\rho_v(\omega,0)$ as a function of $\omega$ at four different temperatures with $\mu=0$ in the upper panel. The peaks in $\rho_v(\omega,0)$ represent the resonance states of the vector meson. It is notable that the locations of the peaks coincide approximately with the radial excited spectrum of the vector meson calculated in Ref. \cite{Fang:2016nfj}. We can see that the peaks in $\rho_v(\omega,0)$ disappear gradually with a width broadening as the temperature increases, until the last one for the lowest lying state becomes flat at some critical temperature, which conforms with the picture that the meson resonance states become unstable and melt gradually with the increase of $T$. We also find that the locations of the peaks shift slightly towards the left with the increase of $T$, which indicates a small mass shift of the in-medium vector meson. The spectral functions $\rho_v(\omega,0)$ at four different chemical potentials with a fixed temperature $T=15\MeV$ are plotted in the lower panel of Fig. \ref{spectr-V}, where we find similar behaviors to those with a fixed $\mu$ which are shown in the upper panel, that is, the in-medium vector meson states melt gradually with the increase of $\mu$.

\begin{figure}
\centering
\includegraphics[width=71mm,clip=true,keepaspectratio=true]{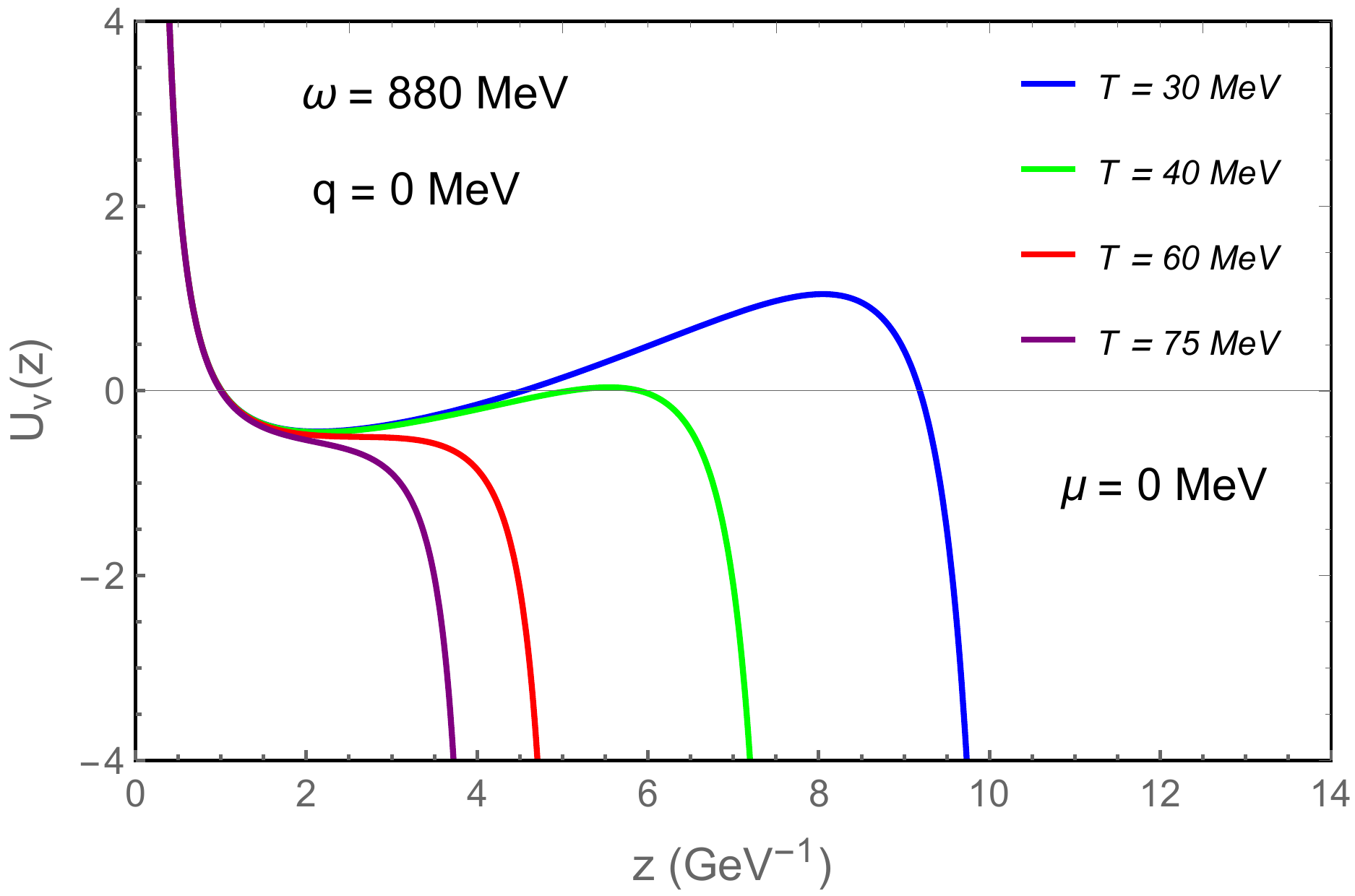}
\caption{The potential $U_v(z)$ for the ground state of the vector meson at four different temperatures in the $\mu=0$ case.} 
\label{U-v}
\end{figure}

We can also look into the in-medium melting properties of the vector meson states from another angle. In terms of the field redefinition $V(z)=e^{\left(\Phi-A\right)/2}f^{-1/2}\tilde{v}(z)$, Eq. (\ref{vec-eom2}) can be recast into the Schr{\"o}dinger form: $\tilde{v}''-U_v(z)\,\tilde{v}=0$ with the potential function
\begin{align}\label{vec-potential}
U_v(z) &=\frac{1}{2}\left(A''-\Phi''\right)+\frac{1}{4}{\left(A'-\Phi'\right)}^{2} +\frac{f'}{2f}\left(A'-\Phi'\right) -\frac{f'^2}{4f^2} +\frac{f''}{2f}-\frac{1}{f}\left(\frac{1}{f}\omega^2-q^2\right),
\end{align}
which should have the convex property to guarantee the existence of bound states for the vector meson. However, with the increase of temperature, $U_v(z)$ loses convexity gradually and eventually becomes monotonic at certain critical temperature $T_c$, as shown in Fig. \ref{U-v}, where we have plotted the curves of $U_v(z)$ at four different temperature $T$ in the case of $\mu=0$ for the lowest lying vector-meson state with mass $\omega=880\MeV$ and $q=0$.

\begin{figure}
\centering
\includegraphics[width=71mm,clip=true,keepaspectratio=true]{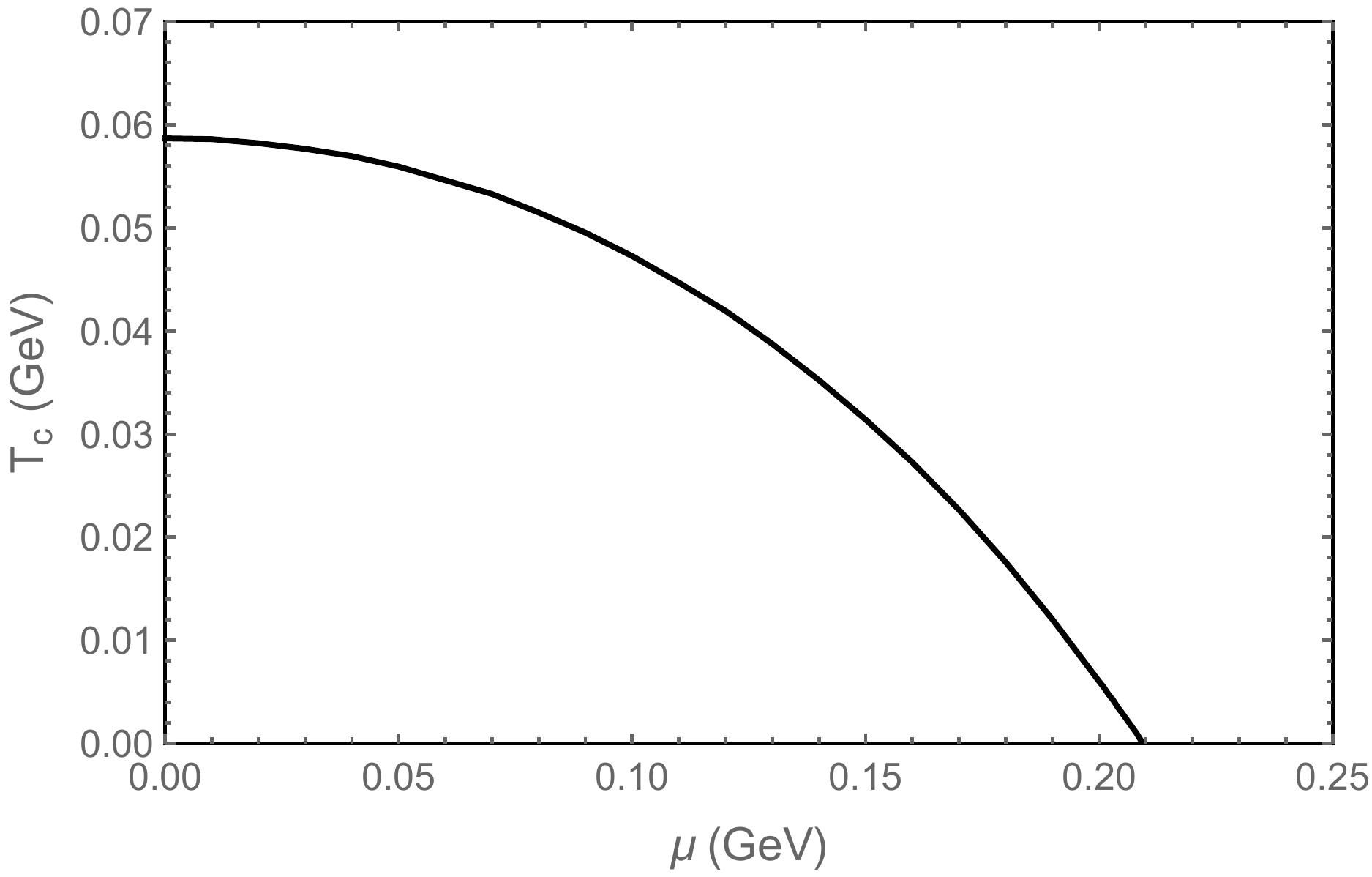}
\caption{The $\mu-T_c$ diagram which is obtained by the monotonicity analysis of the potential $U_v(z)$ in the vector case.} 
\label{fig-T-mu-vec}
\end{figure}

The critical temperature $T_c$ at which $U_v(z)$ becomes monotonic may be defined as the meson melting temperature. We compute $T_c$ at different chemical potential $\mu$ and obtain the $\mu-T_c$ diagram, which is shown in Fig. \ref{fig-T-mu-vec}. It can be seen that the meson melting temperature $T_c$ decreases with the increase of the chemical potential $\mu$, which bears a qualitative resemblance to the behavior of the chiral phase diagram presented in Fig. \ref{fig-T-mu-chi}. However, quantitatively, we find that the meson melting temperature ($T_c\simeq 58\MeV$ at $\mu=0$) is too small compared with the chiral transition temperature $T_{\chi}$ and also the deconfinement temperature indicated by lattice QCD, which is a typical feature of the holographic calculations for the meson melting temperature based on the soft-wall AdS/QCD framework, as has been shown in many other previous studies \cite{Colangelo:2009ra,Bartz:2016ufc}.

\subsection{The axial-vector meson}

Following the same procedure as that in the vector part, we now investigate the thermal spectral properties of the axial-vector meson, whose EOM can be derived from the variation of the action (\ref{V-A-action}) with respect to the transverse component of the axial-vector field $A_{\mu\perp}$ in the $A_z=0$ gauge (we omit the symbol $\perp$ for simplicity),
\begin{align}\label{av-eom1}
\partial_z\left(e^{-\Phi}\sqrt{g}\,g^{zz}g^{\mu\lambda}\partial_{z}A_{\lambda}\right) +e^{-\Phi}\sqrt{g}\,g^{\mu\lambda}g^{\nu\rho}\partial_{\nu}\partial_{\rho}A_{\lambda} 
-g_5^2\,e^{-\Phi}\sqrt{g}\,\chi^2A^{\mu} =0.
\end{align}
Moving to the momentum space with the Fourier transformation $A_{\mu}(x,z) =\int d^4x e^{ip\cdot x}A(p,z)\mathcal{A}^0_{\mu}(p)$ and substituting the metric ansatz (\ref{metric}) into Eq. (\ref{av-eom1}), the EOM of the bulk-to-boundary propagator $A(p,z)$ for the spatial part of the axial-vector field can be obtained as
\begin{align}\label{av-eom2}
\partial_z\left(e^{A -\Phi}f\partial_zA(p,z)\right) +e^{A -\Phi}\left(\frac{1}{f}\,\omega^2 -q^2 \right)A(p,z) -g_5^2\,e^{3A -\Phi}\chi^2 A(p,z) =0.
\end{align}
Similarly, with the replacement of the variables:
\begin{align}\label{av-coord1}
z \rightarrow u\,z_h, \quad   A(p,z) \rightarrow a(u), \qquad 0<u<1,
\end{align}
Eq. (\ref{av-eom2}) can be rewritten in the following form:
\begin{align}\label{eom-a-u-av}
a''(u) +l_1(u) a'(u) +l_0(u) a(u) =0
\end{align}
with $l_1(u)=k_1(u)$ and
\begin{align}\label{eom-a-u-coef}
l_0(u) &=\frac{1}{u^2\left(u^2-1\right)^2\left(1+u^2-Q^2 u^4\right)^2}\Big\{u^2 z_h^2\left[\omega^2 -q^2\left(1 -\left(Q^2 +1\right) u^4 +Q^2 u^6\right)\right]
\nonumber\\ 
&\quad -4\pi^2 (1-u^2)(1+u^2-Q^2u^4)\chi^2\Big\} .
\end{align}

The IR asymptotic solution of Eq. (\ref{eom-a-u-av}) near the horizon has the form:
\begin{align}\label{IR-asy-av}
a(u\sim 1) = \tilde{c}_{+}\,\phi_{+}(u) +\tilde{c}_{-}\,\phi_{-}(u)
\end{align}
with $\phi_{\pm}(u)$ just the same as those in (\ref{in-out-sol}), while the UV asymptotic form of $a(u)$ can also be obtained from Eq. (\ref{eom-a-u-av}) as
\begin{align}\label{UV-asy-av}
a(u\sim 0) &=\tilde{A}\left(1 +\tilde{c}_{l2}u^2\log{u} +\cdots\right) +\tilde{B}\left(u^2 +\cdots \right),
\end{align}
where $\tilde{A}$ and $\tilde{B}$ are two arbitrary constants, and $\tilde{c}_{l2} =\frac{1}{2} z_h^2\left(q^2 -\omega^2 +4\pi^2\zeta^2m_q^2\right)$. We only keep to the $u^2$ term in the asymptotic expansion (\ref{UV-asy-av}). As in the case of vector meson, we take $\tilde{A}=1$ to fix the overall constant of $a(u)$.

To compute the spectral function of the axial-vector meson, we need the retarded Green's function which can be derived as
\begin{align}\label{green-ua}
\tilde{D}^{R}(\omega,q) &=-\frac{C}{z_h^2}\lim_{u\to\epsilon}\left(\frac{1}{u}a^{*}a^{\prime}\right)        \nonumber\\
&=-\frac{2C}{z_h^2}\left[\tilde{B}(\omega, q) +\frac{1}{2}\tilde{c}_{l2} +\tilde{c}_{l2}\log\epsilon\right]
\end{align}
with $C=N_c^2/(64\pi^2 L)$. As aforementioned, this corresponds to the in-falling IR asymptotic solution of $a(u)$ with $\tilde{c}_{+}=0$ in (\ref{IR-asy-av}), which uniquely determines $\tilde{B}(\omega, q)$ at fixed $\omega$ and $q$, and thus the retarded Green's function $\tilde{D}^{R}(\omega,q)$. The spectral function of the axial-vector meson can be obtained by its definition as
\begin{align}
\rho_a(\omega,q) =-\frac{1}{\pi}\,\mathrm{Im}\tilde{D}^R(\omega,q) =\frac{2C}{\pi z_h^2}\,\mathrm{Im}\tilde{B}(\omega, q).
\end{align}

\begin{figure}
\begin{center}
\hskip -0.1 cm
\includegraphics[width=71mm,clip=true,keepaspectratio=true]{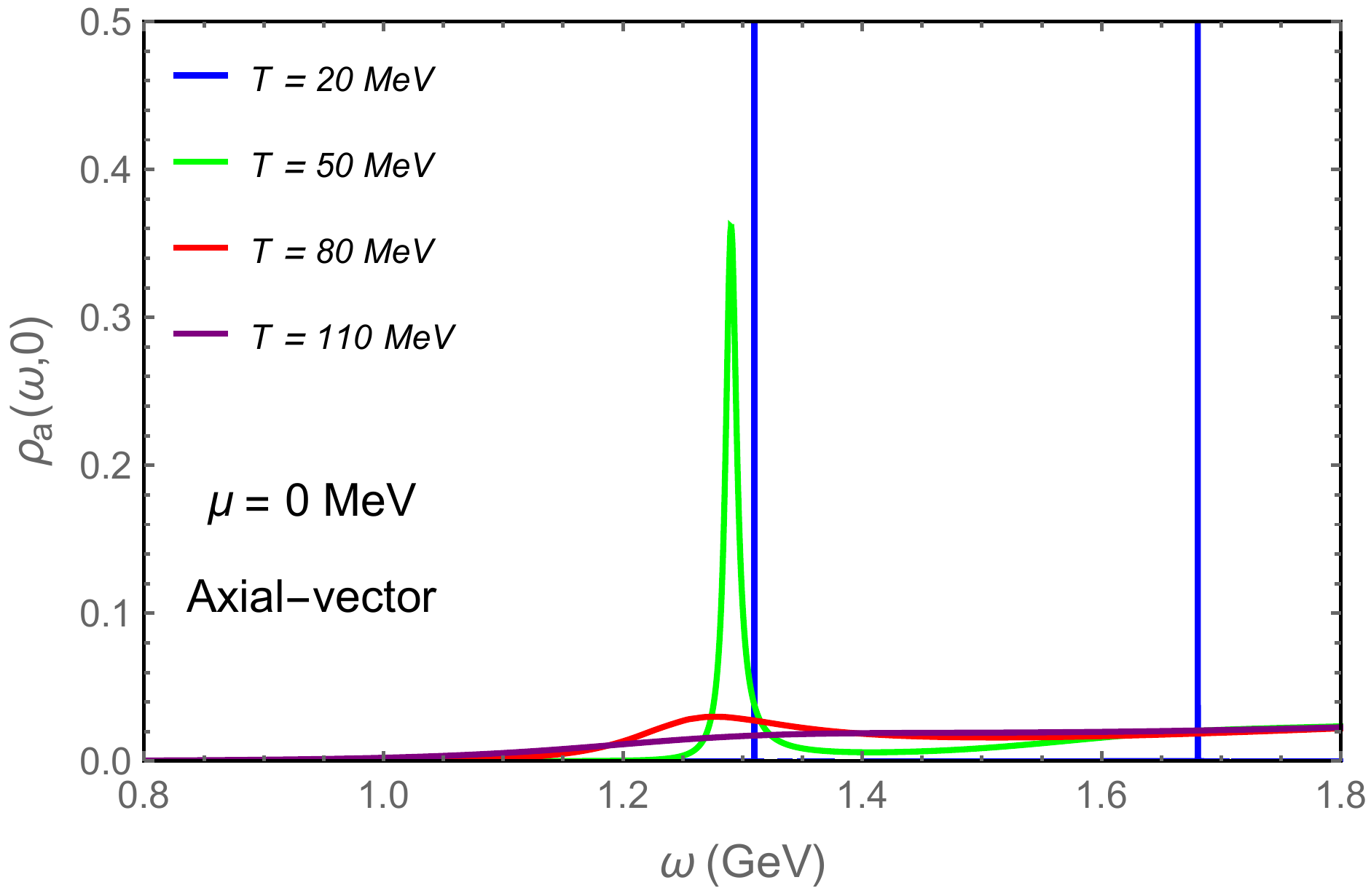}
\vskip 0.3cm  \hskip -0.1 cm
\includegraphics[width=71mm,clip=true,keepaspectratio=true]{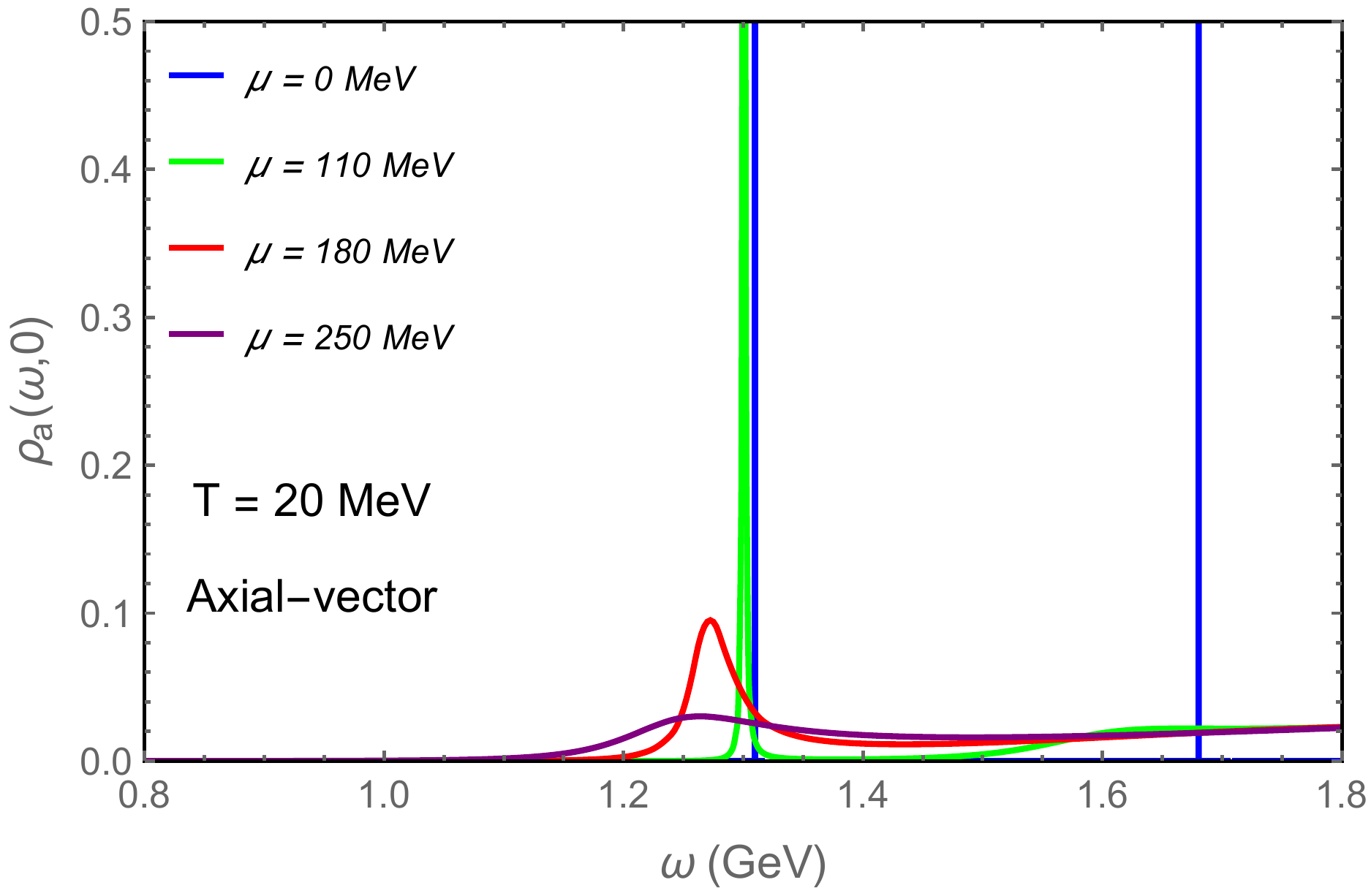}
\vskip -0.8cm \hskip -0.1 cm
\end{center}
\caption{The spectral functions of the axial-vector meson $\rho_a(\omega,0)$. The upper panel shows $\rho_a(\omega,0)$ at four different temperatures in the $\mu=0$ case, while the lower panel shows $\rho_a(\omega,0)$ at four different chemical potentials for a fixed temperature $T=20\MeV$.}
\label{spectr-A}
\end{figure}

We calculate the thermal spectral function of the axial-vector meson $\rho_a(\omega,q=0)$ for four different temperatures at $\mu=0$. The numerical results are shown in the upper panel of Fig. \ref{spectr-A}, where we can see that the axial-vector meson states represented by the peaks in $\rho_a(\omega,q=0)$ melt gradually with the increase of $T$, and eventually the last peak of the ground state of the axial-vector meson disappears at about $T\sim 110\MeV$. We also find that the locations of the peaks shift towards smaller values of $\omega$ with the increase of $T$, as in the vector case. We plot $\rho_a(\omega,0)$ for four different chemical potentials with a fixed temperature ($T=20\MeV$) in the lower panel of Fig. \ref{spectr-A}. We see that the axial-vector meson states melt gradually with the increase of $\mu$, which resembles qualitatively the temperature-dependence of $\rho_a(\omega,0)$ in the $\mu=0$ case, which is shown in the upper panel of Fig. \ref{spectr-A}.

\begin{figure}
\centering
\includegraphics[width=71mm,clip=true,keepaspectratio=true]{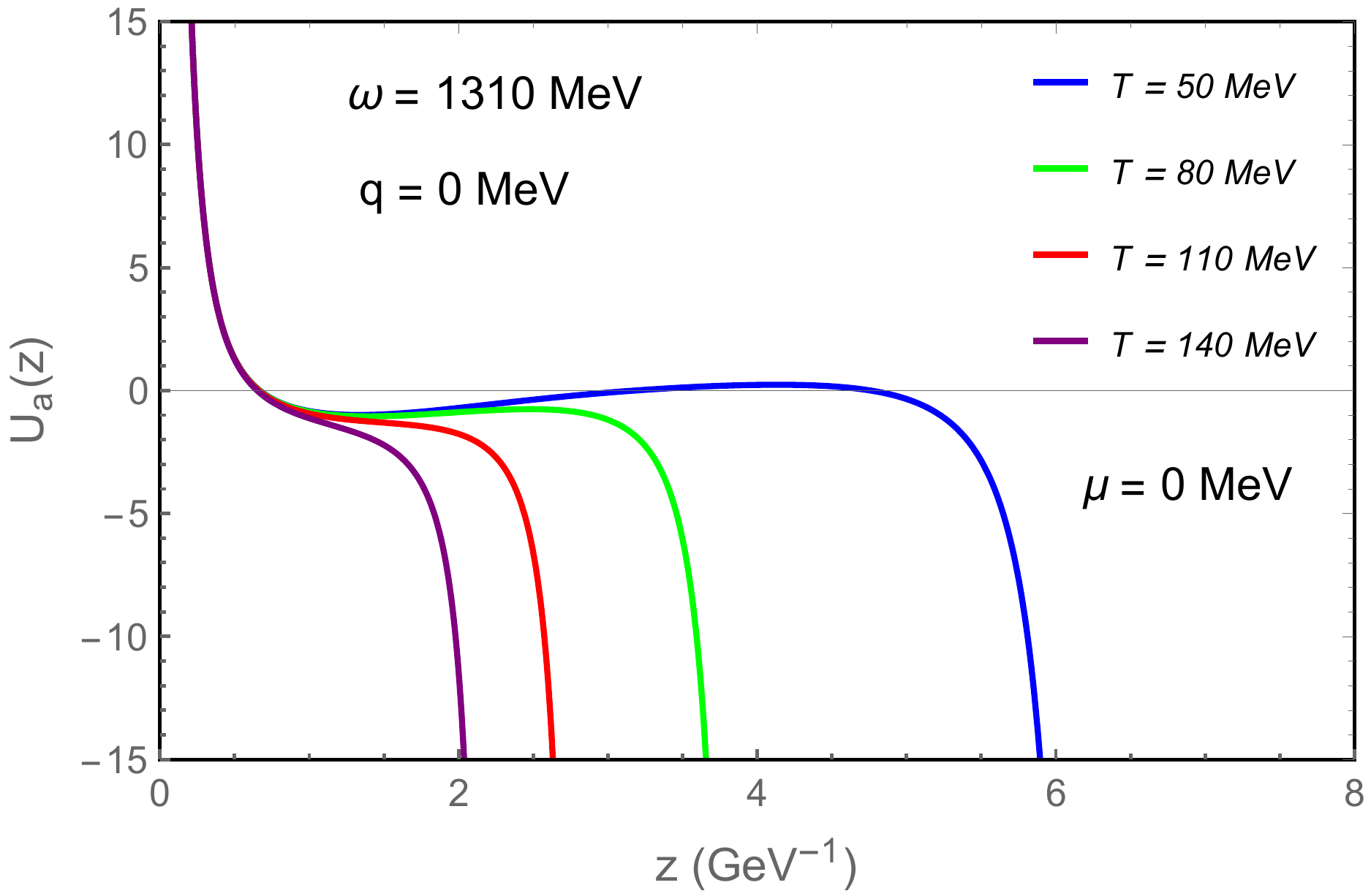}
\caption{The potential $U_a(z)$ for the ground state of the axial-vector meson at four different temperatures in the $\mu=0$ case.} 
\label{U-a}
\end{figure}

As in the vector case, we can also transform Eq. (\ref{av-eom2}) into the Schr{\"o}dinger form: $\tilde{a}'' -U_a(z)\,\tilde{a}=0$ with the field redefinition $A_n(z)=e^{\left(\Phi-A\right)/2}f^{-1/2}\tilde{a}(z)$, and the potential function $U_a(z)$ takes the form
\begin{align}\label{av-potential}
U_a(z) &=\frac{1}{2}\left(A''-\Phi''\right)+\frac{1}{4}{\left(A'-\Phi'\right)}^{2} +\frac{f'}{2f}\left(A'-\Phi'\right)
\nonumber \\
&\quad -\frac{f'^2}{4f^2} +\frac{f''}{2f}-\frac{1}{f}\left(\frac{1}{f}\omega^2-q^2\right) +\frac{g_5^2}{f}e^{2A}\chi^2.
\end{align}
We plot in Fig. \ref{U-a} the curves of $U_a(z)$ for the axial-vector meson with $\omega=1310\MeV$ and $q=0$ for four different temperatures at $\mu=0$. We can see that $U_a(z)$ will lose its convexity with the increase of $T$ and become monotonic at some critical temperature $T_c$. By the same monotonicity analysis as that in the vector case, we know that there will be no bound state of the axial-vector meson when $T>T_c$. Thus the critical temperature $T_c$ can also be defined as the in-medium melting temperature of the axial-vector meson state. We find that $T_c$ for the axial-vector meson is larger than that for the vector meson which is shown in Fig. \ref{fig-T-mu-vec}.

\section{Summary and conclusions}\label{conclusion}

In this work, we give a further study on the improved soft-wall AdS/QCD model with two flavors which is proposed in Ref. \cite{Fang:2016nfj}. The chiral transition behaviors, along with the in-medium meson melting properties, in the case of finite chemical potential have been investigated. We find that the chiral transition with the variation of $\mu$ has the similar behaviors to that with the variation of $T$, i.e., in the chiral limit, it is a second-order phase transition, while in the case of nonzero quark mass it becomes a crossover transition. The $\mu-T$ phase diagram has been obtained by extracting the critical temperature $T_{\chi}$ from the chiral transitions in the chiral limit. We find that the transition curve in the $\mu-T$ plane decreases too slowly when $\mu$ is large enough, which indicates that the improved soft-wall model cannot be applied in the large $\mu$ region. 

We also investigated the varying tendencies of the thermal spectral functions of the vector and axial-vector mesons at different temperatures and chemical potentials. As expected, with the increase of $\mu$ or $T$, the peaks in the spectral function disappear gradually with a small mass decrease of the meson states. Furthermore, we looked into the meson melting properties in terms of the potential functions of the Schr{\"o}dinger-type equations of the mesons. The melting temperature at which the meson states disappear was extracted by the monotonic condition of the potential function, and the $\mu-T$ diagram was obtained for the vector case. We find that the melting temperature at $\mu=0$ in the improved soft-wall model is too small to match with the deconfining or chiral transition temperatures indicated by lattice QCD, which seems to be a common defect for the soft-wall AdS/QCD models \cite{Colangelo:2009ra,Bartz:2016ufc}.

It has been shown in Ref. \cite{Fang:2016nfj} that the improved soft-wall model can give a quantitative description on the light meson spectra and the chiral transition at zero chemical potential. However, we find here that the descriptions on the meson melting properties and the chiral transition behaviors at finite chemical potential are only qualitative. Moreover, this model should be invalid at large enough $\mu$. There may be multiple reasons for the inadequate descriptions by the improved soft-wall model on the low-energy phenomenologies with finite $\mu$. Basically, the AdS/CFT correspondence without string-loop corrections is unable to provide a complete characterization on the low-energy QCD. Nevertheless, the most urgent issue in our case may be that we need to construct a more consistent AdS/QCD model by putting the background part and the flavor part on the same footing. Previous studies have shown that the QCD equation of state and many other thermodynamical quantities can be mimicked by an Einstein-dilaton system or an Einstein-Maxwell-dilaton system in the finite $\mu$ case \cite{Gubser:2008yx,Gubser:2008ny,Gubser:2008sz,DeWolfe:2010he,Gursoy:2007cb,Gursoy:2007er,Gursoy:2008bu,Gursoy:2008za,Rougemont:2015wca}. Thus, we shall consider such an dynamical background system in order to give a more realistic description for the chiral transition and the in-medium properties of hadrons.

\section*{Acknowledgements}
This work is supported by the National Natural Science Foundation of China (NSFC) under Grant No. 11905055 and the Fundamental Research Funds for the Central Universities under Grant No. 531118010198.

\bibliography{refs-AdSQCD}   


\end{document}